\documentclass[twoside,11pt]{article}
\usepackage{jair, theapa}
\usepackage{rawfonts}
\usepackage{algorithm}
\usepackage{algpseudocode} 
\usepackage{amssymb}
\usepackage{enumerate}
\usepackage{graphicx}
\usepackage{psfrag}
\usepackage{tikz}
\usepackage{multirow}
\usepackage{url}

\ShortHeadings{Simulation and Analysis of Container Freight Train Operations at Port Botany}
{Guimarans, Harabor and Van Hentenryck}
\firstpageno{1}

\begin{document}

\title{Simulation and Analysis of \\ Container Freight Train Operations at Port Botany}
\author{\name Daniel Guimarans \email d.guimarans.serrano@hva.nl \\
\addr Amsterdam University of Applied Sciences \\
\addr NICTA Optimisation Research Group \\
\AND
\name Daniel Harabor \email daniel.harabor@nicta.com.au \\
\addr NICTA Optimisation Research Group \\
\AND
\name Pascal Van Hentenryck \email pvanhent@umich.edu\\
\addr University of Michigan 
}

\maketitle

\begin{abstract}
  Over two million containers crossed the docks at Sydney's Port Botany in
  2011/12; a figure that is forecast to increase more than threefold by the
  end of the next decade.  To cope with such large growth in volumes, the New
  South Wales (NSW) State Government plans to double rail mode share at the port by
  the year 2020.  Conventional wisdom from industry and the media says that
  existing infrastructure cannot handle such volumes.  In this paper, we use a
  combination of data analytics and simulation to examine operations at the
  port and evaluate the efficacy of current infrastructure to handle projected
  growth in volumes.  Contrary to conventional wisdom, our findings indicate
  that current rail resources appear distinctly under-utilised. Moreover: (i)
  the peak rail capacity of Port Botany is 1.78 million TEU per annum, over
  six times higher than 2011/12 rail volumes; (ii) there are no
  infrastructural impediments to the achievement of peak rail capacity; (iii)
  operational changes, not infrastructural investment, are the key to
  unlocking the potential of the port; (iv) Port Botany is well positioned to
  handle projected increases in container volumes over the next decade and
  beyond, including the 28\% rail mode share target established by the NSW
  Government.


\end{abstract}


\section{Introduction}
\label{sec::introduction}

Port Botany is Australia's second largest container port, handling
approximately one third of the nation's maritime container traffic.
In 2011-12, total volumes at the port exceeded two million twenty-foot
equivalent units (TEU), with 86\% of all containers transported by
road and the remaining 14\% transported by rail.  Container volumes
are expected to increase annually over the next decade and projected
to reach 3.6 million TEU by 2020, $\sim 5$ million TEU by 2025 and
over 7 million by 2031~\cite{nsw13}.  The New South Wales (NSW) State
Government is concerned that future growth at the port will result in
large numbers of additional trucks on Sydney's already congested
roads. To handle the problem, the government has established a 28\%
rail mode share target for container freight at Port Botany by the end
of the decade~\cite{nsw2021}.

The best way to achieve this goal is a contentious subject that has generated
much discussion.  For example, according to one recent analysis from industry
media~\cite{cameron14}, ``Not only is rail capacity insufficient for current
container demand, there is no rail capacity to meet future container demand.
[...] Port Botany's effectiveness is already impacted by inadequate road and
rail infrastructure''.

In previous studies, commissioned by government and supported by industry, it
has been suggested that rail at Port Botany has limited opportunities to
benefit from economies of scale and that key components of its rail
infrastructure are limiting factors. There are three widely held but largely
untested perceptions arising from these studies:
\begin{enumerate}
\item Rail operations are bottlenecked by a 3km section of single track that
provides access to and from the port~\cite{ipart-rpblt-08,aecom12}.
\item The configuration of the DP World rail terminal is an
impediment to increased rail volumes~\cite{ipart-rpblt-08,aecom12}.
\item Port Botany would be better served in the future by consolidating rail
operations at a new and centralised off-dock rail terminal~\cite{aecom12}.
\end{enumerate}

Addressing any of these issues requires large-scale investment in new
infrastructure. For example, the asked-for single-line track
duplication has been estimated to cost \$210M
AUD~\cite{NSWParl11}. Given the scale of the expenditure, it is
important to understand precisely how these different options compare,
how they should be prioritised and when they should be enacted. In
this paper, we aim to undertake a principled analysis of the factors
that impede current rail volumes at Port Botany. Along the way, we
also endeavour to test the veracity of each of these perceptions.


In the first instance, we undertake a data-driven analysis of current
rail performance using six (6) months of operational data from the
period September 2012 to February 2013. We study port performance
using a range of metrics including (i) train timeliness; (ii) train
utilisation; (iii) terminal utilisation; and (iv) yard congestion.  We
find that, contrary to popular beliefs from industry and from the
media (e.g.~\cite{aecom12,cameron14}), rail resources at Port Botany
appear distinctly under-utilised. Moreover, the location and
configuration of existing rail terminals is not an impediment to
increased rail volumes.

\begin{figure}[tb]
       \begin{center}
		   \includegraphics[width=\columnwidth]
			{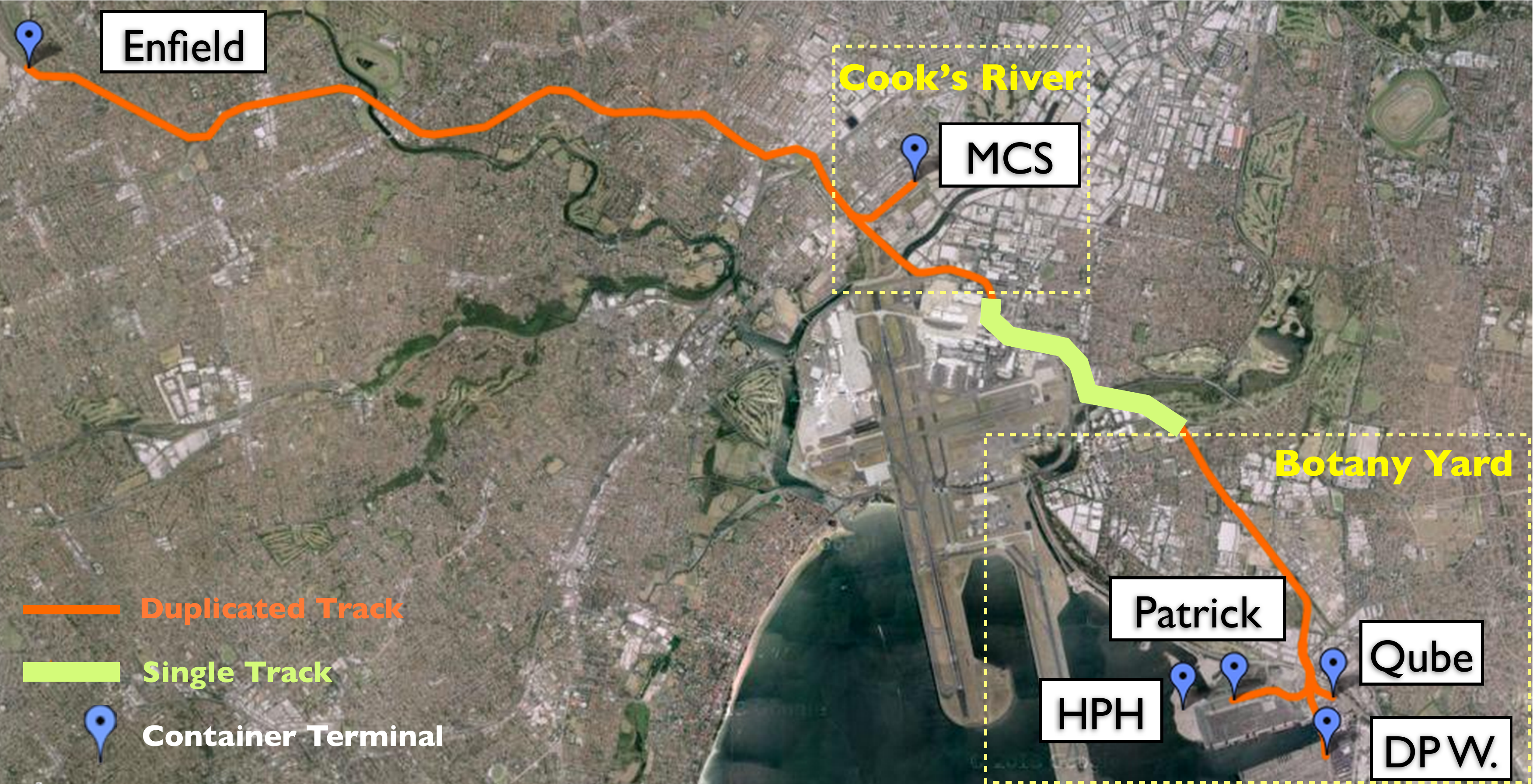}
       \end{center}
       \vspace{-2em}
       \caption
{
A map showing the extent of the Enfield-Botany rail corridor. Labelled
are the locations of the four current rail terminals (DP World, MCS, 
Patrick and Qube) and the approximate locations of two new rail terminals 
due to commence operations in 2014 (Enfield ILC and HPH).
}
\label{fig::enf_by_map}
\end{figure}

Next, we construct a simulation-based model of Port Botany and measure peak
rail capacity (in terms of container volumes) in a range of infrastructural
case-studies. Our simulation focuses on a 20km ``last-mile'' rail corridor
between the Sydney suburb of Enfield and Port Botany
(Figure~\ref{fig::enf_by_map}).  The model we create is at the level of
individual rail sections and sidings and includes detailed servicing
operations at container terminals inside the corridor. Using this model we
show that the current rail capacity of Port Botany is over 1.7M TEU; i.e.
well above the levels needed to meet the NSW Government's 28\% rail mode share
target over the medium term (cf. 1M TEU by 2020 and 1.4M TEU by 2025).
These figures can be achieved without any investment in new infrastructure
such as a centralised off-dock rail terminal or the planned duplication of 3km
of rail track.

Our findings show that the key to unlocking the potential of rail at
Port Botany lies not in infrastructural upgrades but rather
operational changes such as dynamic train scheduling, flexible
servicing windows, improved staging practices, and ``dedicated'' train
services that visit one stevedore per trip.
\section{Methodology}
\label{sec::methodology}
We give a brief description of the main datasets used in 
for the analytical part of our study and an overview of 
the simulation model we employ thereafter.

\subsection{Data Sources}
\label{sec::data}

Our analysis and results are informed by a range of primary and secondary 
data sources. The largest of these are two databases
provided to us by Sydney Ports Corporation:
\begin{itemize}
\item A database of Daily Operations Plans (DOPs) which are compiled
  by Australian Rail Track Corporation (ARTC).  This database contains
  information for 1992 separate freight services that operated at
  Botany Yard in the target six month period.  Each DOP records
  details such as planned and actual arrival times, planned and actual
  departure times, planned and actual servicing times, and the point
  of origin and point of return for each train.

\item A database of operational performance measures compiled by rail crews
working at waterfront terminals.  This database contains operational 
performance metrics for 1324 separate services (938 trains) across the
target six month period. Recorded details include the scheduled number of
lifts performed during a servicing operation, the actual number of lifts,
container type (import, export) and size (20' or 40') and the total
lifting time for all containers.
\end{itemize}

In addition, we collected and analysed a wide range of supplemental
data sources. Such data includes details of operational practices and
constraints gathered during interviews with members of the port
community (terminal operators, rail operators, rail network owners and
others) and direct observations of train operations which were taken
during several visits to Port Botany.

\subsection{Simulator Overview}
\label{sec::methodology::sim}

Much of this study is based on analysis of data from a discrete-event
simulator that models rail operations inside the 20km dedicated freight rail
corridor between Port Botany and Enfield.  The purpose of
our simulator is twofold: (i) to reproduce current rail operations at Port
Botany and thus better understand the factors which impede rail productivity;
(ii) to establish the peak rail capacity of Port Botany in a range of
infrastructural case-studies.

The rail corridor, which is the focus of our simulation, connects Port
Botany to the Sydney Metropolitan Freight Network. It is used
primarily by container-carrying freight trains and is home to four
(since 2015, five) intermodal rail terminals. Two of these terminals
(since 2015, three) are on-dock stevedore terminals and two are empty
container parks. With the exception of one 3km section, the system is
comprised entirely of duplicated track, some of which is configured
for bi-directional running. For further details of the physical
infrastructure and operational procedures and constraints (See
Section~\ref{sec::operations}).

\subsection{Simulator Design}
Our simulator is constructed in the Java programming language and makes use 
of the freely available simulation library DESMO-J~\cite{lechler1999desmo}
(we use version 2.3.5, available from~\url{http://desmoj.sourceforge.net}).
Figure~\ref{fig::simsys} gives a high level overview of the system. We briefly
discuss each of its main components:

\begin{figure}[tb]
    \begin{center}
      \includegraphics[width=\textwidth]{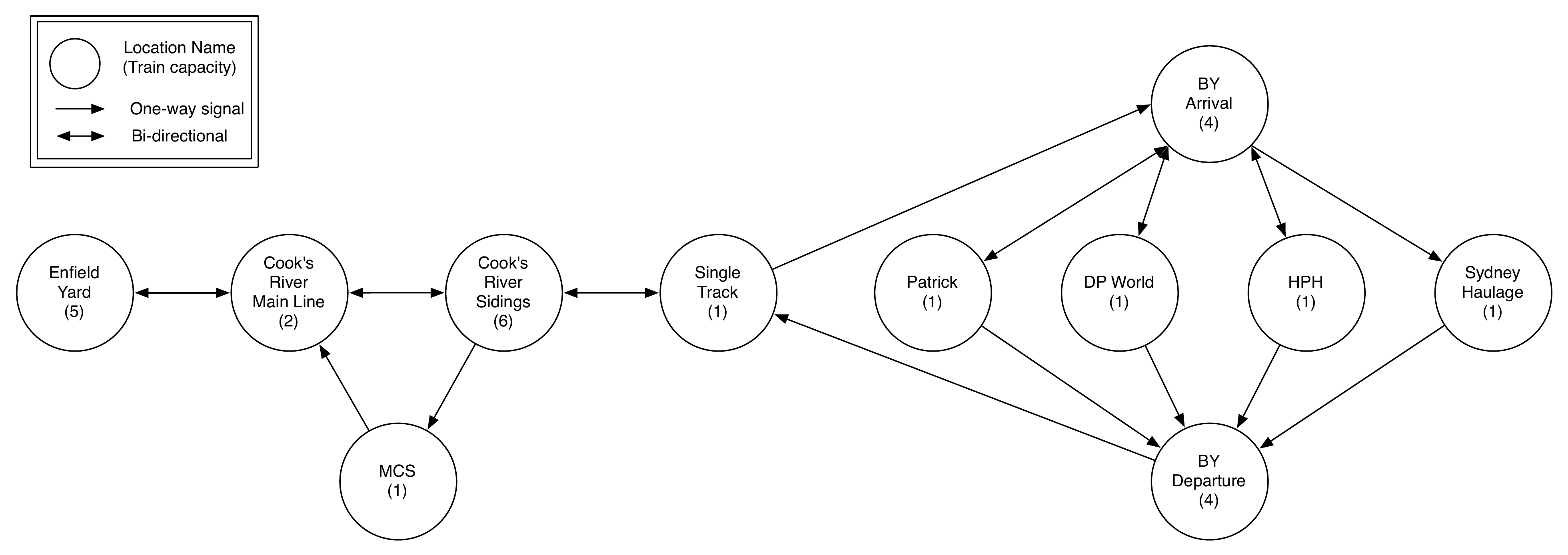}
    \end{center}
    \vspace{-1em}
    \caption{
    An overview of the architecture of our simulation model. Queues
    are represented by nodes. Their capacity is indicated in parentheses. 
    Edges represent rail connections.
        }
    \label{fig::simsys}
\end{figure}

\begin{itemize}

\item The source and sink of the system is located at the Enfield
  Marshalling Yard. Trains originate at Enfield according to a
  pre-determined distribution or according to a fixed (e.g.,
  historical) schedule.

\item There are 5 intermodal rail terminals in the system, each having one or
more available sidings where trains can be shunted, split (if necessary) and
subsequently serviced.  We model each terminal as a queue with capacity equal
to one (1).

\item Two sets of sidings, collectively known as Botany Yard and Cook's
River, are used for staging trains before and after servicing. 
We model each of these yards as single queues with capacity equal to the number
of physical sidings.

\item A 3km section of single track connects Cook's River with Botany Yard.
We model this section of track as a single queue with capacity equal to one (1).
\end{itemize}

\subsection{Simulator Inputs}

To simulate train operations, we use a variety of empirical
distributions computed from operational datasets described in
Section~\ref{sec::data}.  These distributions are used to determine:
(i) stevedore lift rates; (ii) shunt in and shunt out time (i.e., the
time required to propel trains both into and out of each terminal);
(iii) variance between planned and actual number of containers carried
on a train; (iv) possible delays between placement of a train and
starting the service. Other parameters include headway time between
trains and the period of time which is simulated (typically 365 days).

Each parameter can be adjusted. This flexibility allows us to simulate
different scenarios and study, e.g., the impact of higher lift rates or peak
capacity under ideal shunting operations. In the same way we can also add or
remove infrastructure (e.g., add or omit a terminal) and modify the capacity of
infrastructure (e.g., to simulate the effect of single-line duplication, the
addition extra sidings in the corridor, or the impact of parallel train
servicing operations).

\subsection{Simulator Outputs}

The output from the simulator is in the form of train attributes,
train schedules, and stevedore performance data. In the case of
trains, we record information such as the length of the train, the
number of wagons, the number of planned containers, and the number of
actual containers. We also generate a detailed record of train
movements within the system and the times they occur (e.g., arrival to
the Enfield Marshalling Yard, entry/exit at a stevedore terminal,
lifting start/finish, etc.).  In the case of terminals, we compute and
record average, minimum and maximum servicing times, as well as number
of trains serviced. We also calculate total servicing time and
terminal utilisation. All queues in our model are traced, providing
detailed logs and summary figures. Thus, additional information about
infrastructure usage can also be obtained from the simulation, e.g.,
single line utilisation.

\subsection{Simulator Visualisation}

\begin{figure}[t!]
    \begin{center}
      \includegraphics[scale=0.43]{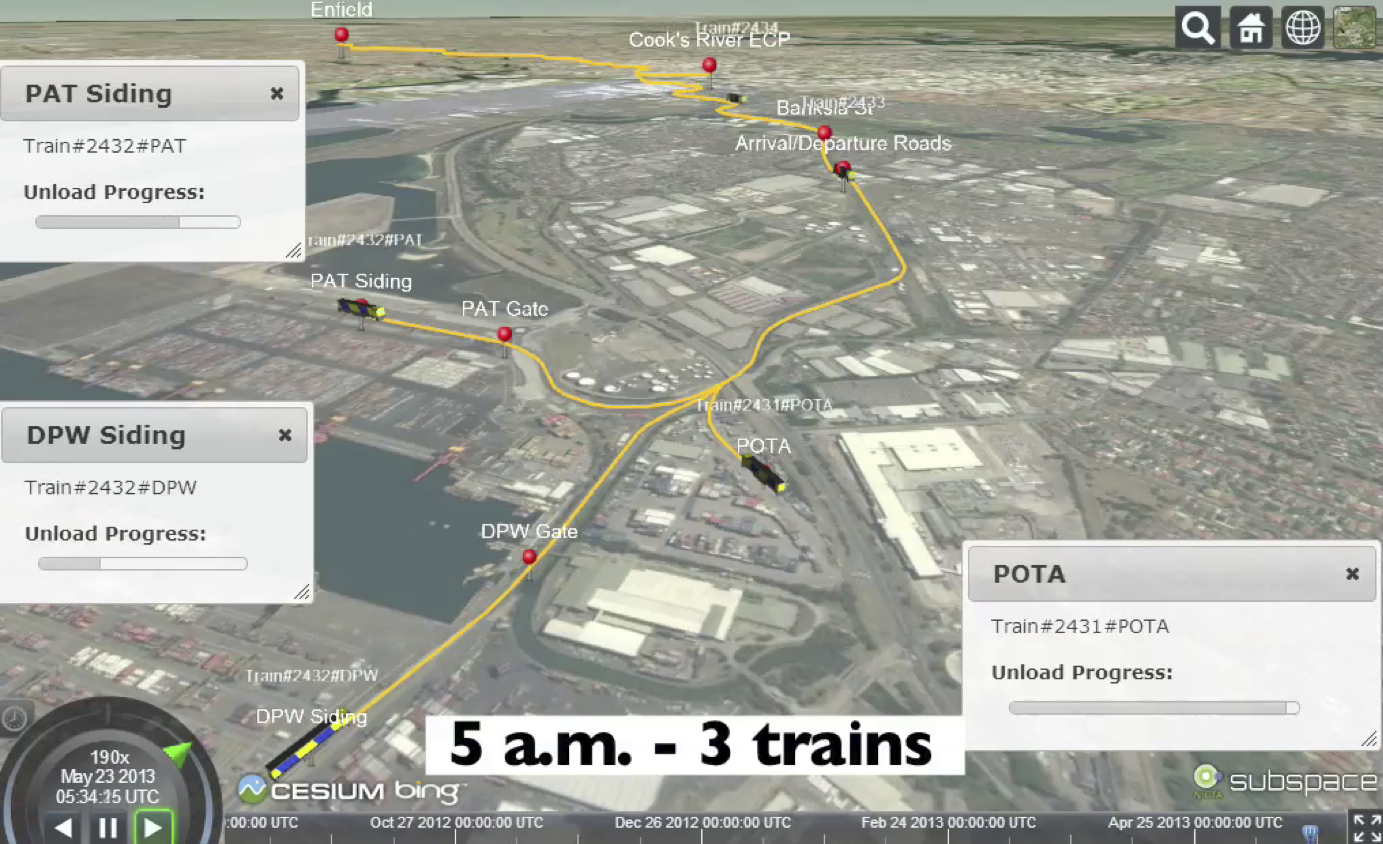}
    \end{center}
    \caption{The visualisaton component of our simulator.}
    \label{fig::simvis}
\end{figure}

As part of this work, we have also developed a visualisation component
that allows us to more easily communicate the results of
simulation-based analysis to our client stakeholders (Sydney Ports
Corporation) and the members of the port community more generally.
The visualiser is built on top of NICTA SubSpace; a freely available 
geospatial visualisation library (\url{http://subspace.nicta.com.au}).
It takes as input a rail network map and a detailed train
schedule. The output is an animation that shows trains moving through
the system and being serviced at various terminals along the way.  An
example of this animation is shown in Figure~\ref{fig::simvis}.

\section{An Overview of Rail Operations at Port Botany}
\label{sec::operations}

The Enfield-Botany rail corridor (shown in Figure~\ref{fig::enf_by_map}) is an
approximately 20km section of (mostly duplicated) rail track connecting the
Enfield Marshalling Yard and Port Botany. The corridor is dedicated freight
and primarily used by trains carrying containers to and from intermodal
terminals located in and around the port precinct. These terminals are:
\begin{itemize}
\item Patrick, DP World, and Hutchinson (HPH). These waterfront terminals 
facilitate the transfer of containers between ships and rail and road transport.
They operate 24/7 with HPH due to commence rail servicing operations in 2015.
\item Sydney Haulage and MCS. These are empty container parks operated 
respectively by Qube Logistics and Martime Container Services.
These terminals are used for storing and providing empty containers for lease and export, 
for unloading trains that have missed their stevedore windows, for refuelling services and
for staging trains that are waiting for an outbound path.
\item Enfield. This inland intermodal terminal, operated by Hutchinson, 
is planned to commence operations in the corridor in the near future. 
\end{itemize}

Another important part of the rail infrastructure is Botany Yard. This collection of
rail tracks and sidings, approximately 3km in length, is used to stage
trains operating at waterfront terminals.  The yard is divided up into 2
arrival roads and 2 departure roads. It can accommodate up to 8 trains of
length 650m or less (2 per arrival and departure road) or 4 trains of length
longer than 650m.  Access to Botany Yard is by way of a 3km section of
single-track (i.e. non-duplicated) between Mascot (near MCS) and the waterfront.


\subsection{Train Operations}
\label{sec::current::trains}
There are three categories of container trains operating in the Enfield-Botany rail corridor.
\begin{enumerate}
\item \emph{Dedicated trains} that only visit one stevedore terminal (and possibly Sydney Haulage);
\item \emph{Split trains} that visit both stevedore terminals (and possibly Sydney Haulage); 
\item \emph{Non-stevedore trains} that only visit Sydney Haulage or MCS.
\end{enumerate}

Each train consists of one or more locomotives and a collection of
wagons called a \emph{rake}.  When a train arrives into Botany Yard,
the locomotive moves to the rear of the rake. This operation, known as
a \emph{run-around}, simplifies servicing operations
and 
must occur before the train can enter any of the terminals.

Once the locomotive is positioned at the rear of the rake, the train
waits in the yard until it is \emph{called-up} by the terminal
operator for servicing.  The train is \emph{shunted} (i.e., propels)
into the appropriate \emph{siding} and the wagons that are to be
serviced are disconnected from the remainder of the rake. At this
point, the train is said to be \emph{placed} and servicing can
begin. During servicing the remainder of the train (possibly just the
locomotive) can either wait outside the entrance to the terminal or
propel into another terminal.

Once servicing is finished, the train is \emph{called-out} by the
terminal operator.  At this point the locomotive or the remainder of
the train returns to pick up and re-attach the serviced wagons, the
crew perform a safety inspection of the entire rake and the train
\emph{exits} (i.e. propels out) from the terminal and into the yard to
wait; either for its next call-up or for its scheduled departure time
from the yard.

\subsection{Waterfront Operations}
\label{sec::current::stevedore}

Each rail terminal offers pre-allocated and ad-hoc time windows for train
servicing operations. In order to reduce risk and minimise delays, rail
operators will pre-allocate windows; sometimes up to months in advance.  A
``standard'' window is 90 minutes in length with 30 minutes of this time
allocated for shunting the train in and out of the terminal and the remaining
60 minutes allocated for lifting containers on and off the train.

\section{Current Performance (Data Analytics)}
\label{sec::cperf}
In this section, we evaluate the performance of container-freight
train operations in the Enfield-Botany rail corridor. We analyse a six
month period spanning September 2012 to February 2013, all inclusive.
During this period, a total of 1992 container-freight trains visited
Port Botany. We analyse the makeup and performance of these trains,
the performance of servicing operations at each of the two stevedore
terminals and we look at the utilisation of Botany Yard.
Additionally, we investigate three popular perceptions, originating
within the port community, regarding rail operations at Port Botany:
(i) the perception that Botany Yard is congested; (ii) the perception
that the location and configuration of some current rail terminals is
an impediment to increased rail volumes; (iii) the perception that the
3km single-track section to access the port precinct constrains rail
volumes.  In each case, we find that these perceptions are not
accurate.

\subsection{Perception 1: Port Botany has inadequate rail infrastructure}
The perception that Port Botany does not have adequate rail infrastructure is
one that has appeared in our discussions with members of the port community.
It has also appeared in recent analysis by industry media~\cite{cameron14}.
We test this perception in two ways: (i) by measuring the current utilisation
of available rail sidings, and thus congestion, in Botany Yard; (ii) by
analysing the current utilisation of stevedore terminals.

\paragraph{Yard Congestion}
Botany Yard can accommodate up to 8$\times$650m trains in total across its two
arrival roads and two departure roads.  In addition, three more trains can be
serviced in the yard -- one at each of the two waterfront terminals and one at
Sydney Haulage. The current capacity of Botany Yard is therefore
11$\times$650m trains. We compare this maximum figure with train volumes
observed during the target six month period.

\begin{figure}[tb]
\begin{center}
\includegraphics[width=0.95\columnwidth]
{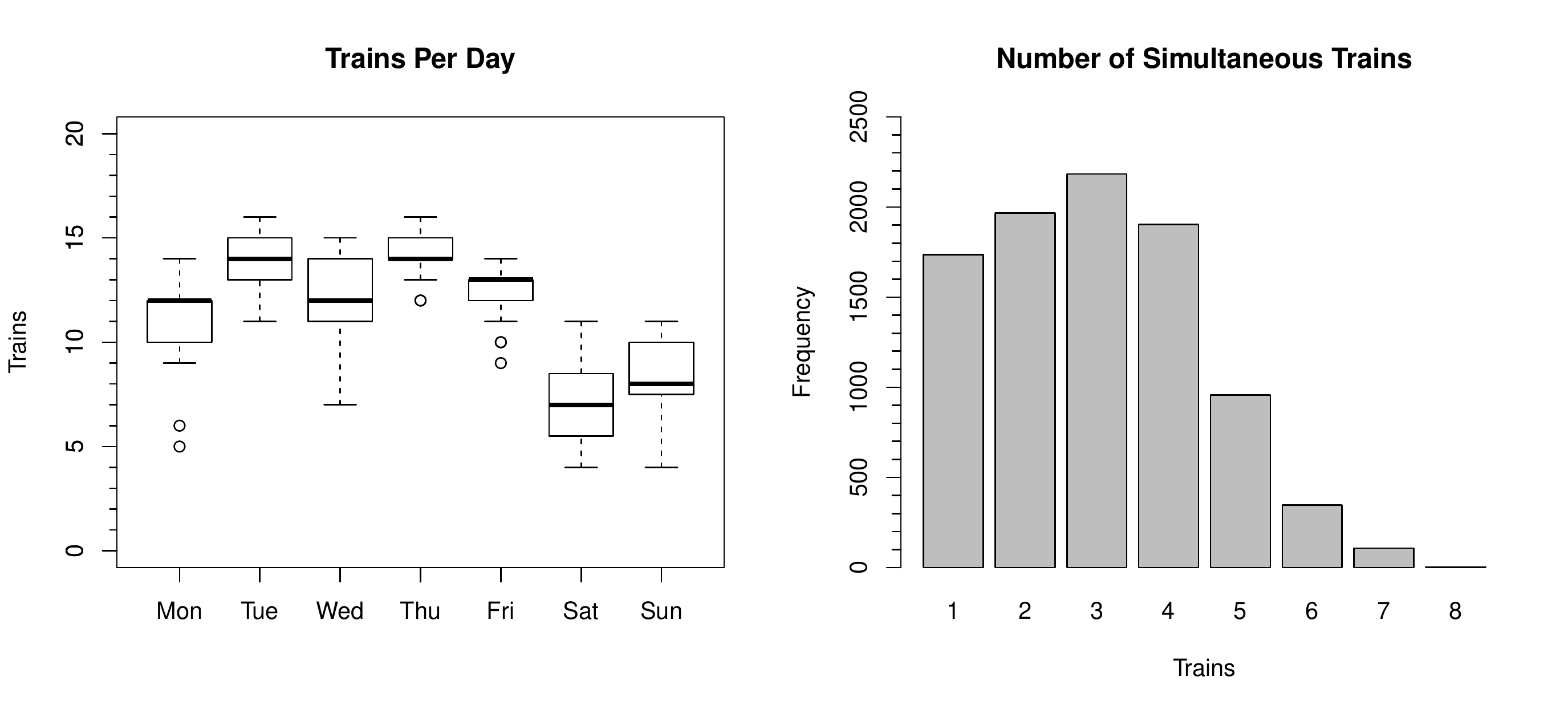}
\end{center}
\vspace{-3em}
\caption
{
Distribution of daily arrivals in Botany Yard by day-of-the-week (left)
and distribution of the number of trains operating simultaneously
in Botany Yard (right).
}
\label{fig::cperf::congestion}
\end{figure}

We begin with Figure~\ref{fig::cperf::congestion} (left) which shows
that, on an exceptionally busy day, up to 16 trains can arrive at the
port during a 24 hour period. The expected figure is less; between 11
to 14 arrivals per weekday and 7 to 8 on weekends.  In
Figure~\ref{fig::cperf::congestion} (right), we show a frequency count
for the number of simultaneous trains operating in the yard at any one
time.  To compute this measure we count how many trains are in the
yard each time the ARTC's DOP indicates that a train movement has
occurred. A movement can be a train arrival or departure, or a train
entering or exiting a terminal.

We find that in the vast majority of cases, there are only between 1-6
trains operating simultaneously in the yard. This suggests that the
current utilisation of rail track resources in Botany yard is between
9-55\%. The analysis is similar if we account for the length of each
train and not just the total number. We find that, in 89\% of cases,
there are no more than 6$\times$650m sidings in use at any one time.
Although trains of length greater than 650m account for 17\% of all
rail traffic, they do not often operate in the yard in large numbers
at the same time.

\begin{figure}[tb]
\begin{center}
\includegraphics[width=0.95\columnwidth]
{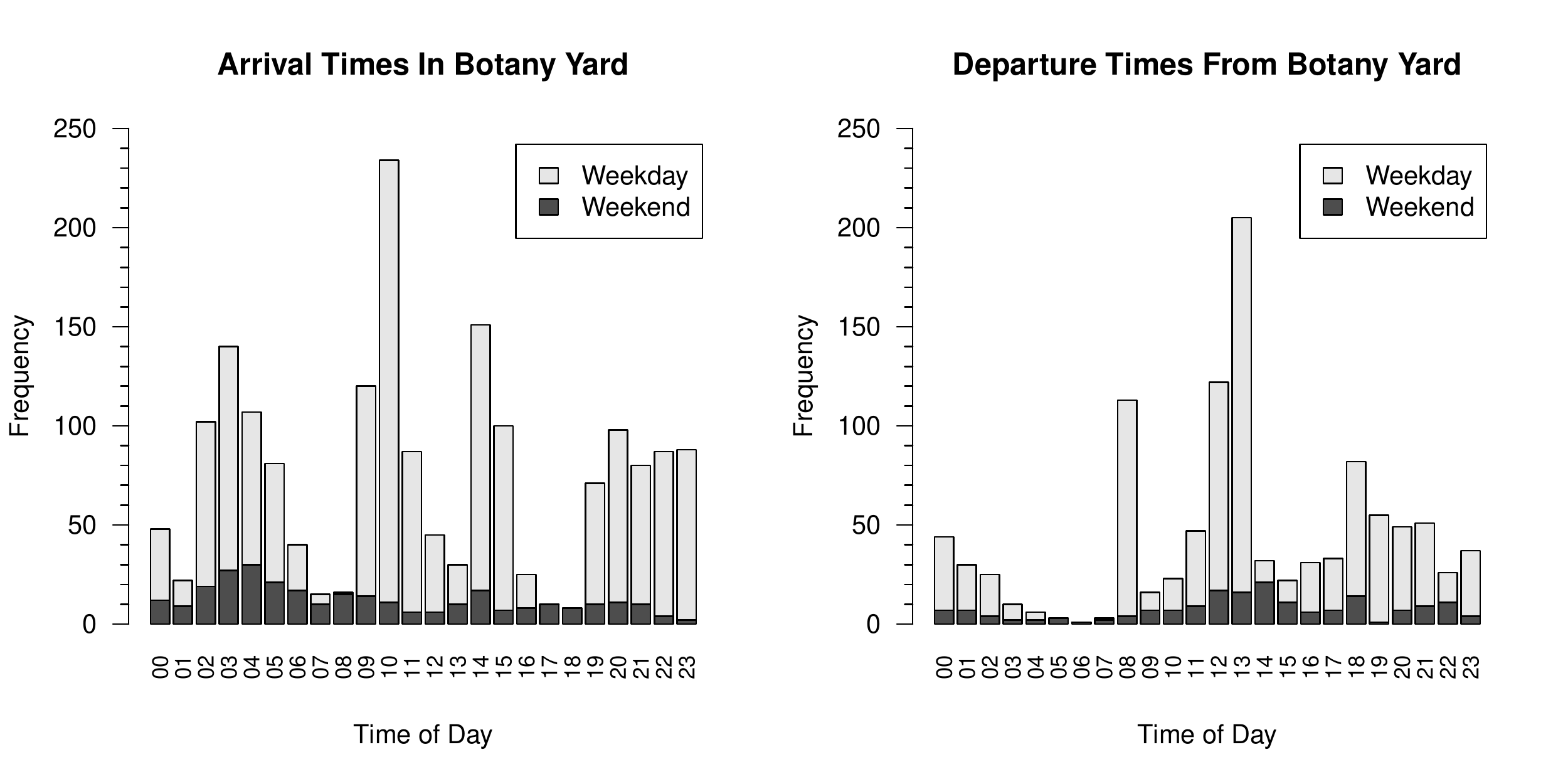}
\end{center}
\vspace{-3em}
\caption
{
Frequency plot of arrival times (left) and departure times (right) for
trains operating at Botany Yard. Note that our available data 
only indicates a departure time $\sim$71\% of all cases.
This means we count fewer departures than arrivals.
}
\label{fig::cperf::arrdep}
\end{figure}

Our analysis thus far suggests that Botany Yard is not a source of
congestion for rail traffic.  However this finding appears contrary to
perceptions held by some members of the port community.  For an
alternative perspective consider Figure~\ref{fig::cperf::arrdep}.
Here we give the frequency of arrival and departure times for trains
operating in the yard.  We can see that there are significant peaks
during the day, often around the start and end of the freight curfew
on Sydney's passenger network.  Depending on specific arrival,
departure, and servicing times, it is entirely plausible that trains
can incur delays. e.g., due to simultaneous demand for shunting
operations at the terminals. We will come back to this issue later on.

\paragraph{Terminal Utilisation}
Next, we analyse the perception that Port Botany has inadequate rail
infrastucture from the point of view of available resources at
waterfront terminals.  We distinguish between two activities: lifting
and waiting. Lifting is the time between the first and last lift, as
reported by stevedores. Waiting is the amount of servicing time that
remains and includes both shunting time and idle time (e.g., waiting
for the locomotive to return).  We omit detailed performance data for
reasons of confidentiality.

During the six month study period a total of 2348 servicing operations
were carried out by stevedores. In one terminal trains are serviced at
a (near) constant lift rate, largely independent of the number of
containers the train is expected to carry.  Waiting times at this
terminal are between 20-35\%, most of which is shunting (i.e., during
servicing operations the terminal appears well utilised).  At the
other terminal the number of lifts per hour is highly variable and
increases or decreases with the number of expected containers. Waiting
times are between 30-60\% (i.e.  during servicing this terminal
appears under-utilised).

For a clearer picture, we also compute the total proportion of time
that each terminal spent servicing a train (i.e. lifting or waiting)
during the study period.  Using this measure the two terminals have a
utilisation rate of 43\% and 66\%; i.e. both appear under-utilised.

\subsection{Perception 2: Rail sidings at the DP World terminal are an impediment}

The DP World rail terminal is configured with $3\times350m$ sidings.
Trains longer than 350m require more complicated shunting operations
when accessing this terminal as the train needs to be split across two
or more sidings prior to servicing. This situation has led some within
the port community to conjecture that a bottleneck exists within the
terminal \cite{ipart-rpblt-08,aecom12}.

To investigate this perception, we analysed six months of DOP data,
recorded by ARTC. The data indicates when a train moves into and out
of each stevedore terminal.  On first analysis, we find that shunting
into and out of the DP World terminal can indeed take 4-5 minutes
longer than shunting into and out of the Patrick terminal. Further
investigation however reveals the issue is more complex.

As part of our study, we engaged directly with railway engineers from
ARTC and undertook on-site visits to their monitoring facility at Port
Botany.  These discussions reveal that shunt in and shunt out times
recorded by ARTC correspond to the times when a train leaves the track
circuits and when it re-appears on the track circuits (railway tracks
at Botany Yard are instrumented with sensors and the location of each
train is precisely monitored by ARTC).  At DP World, the track
circuits end approximately 300m before the stevedore gate whereas at
Patrick the circuits end very close to its gate.  A typical port train
takes 1 minute to traverse 100m of siding. If we adjust (i.e.,
subtract) this travel, the shunt time difference between the two
terminals vanishes almost entirely.  This finding is consistent with
direct observations that we made while visiting both rail terminals .
In particular, we have seen that splitting at DP World can be very
fast.  The crossover track is located next to the gate and the train
can propel quickly from one siding to the other. We measured the
overhead at one minute, which is consistent with our data analysis.

We conclude that there is no appreciable difference between shunting in and
shunting out at either terminal despite the additional splitting operation at
the DP World terminal.  We believe this perception stems from the fact that,
in practice, shunting time and lifting time might be similar for trains
carrying only few containers.

\subsection{Perception 3: The 3km single-track accessing Port Botany is a bottleneck}
A 3km section of single-track connects Botany Yard to the dedicated freight line
from Enfield. This contentious piece of infrastructure is often identified
as a bottleneck and an impediment to the growth of rail~\cite{ipart-rpblt-08,aecom12}.

ARTC (the rail network owners), estimate the maximum capacity of this
line at 36 return train-trips per day, on average \cite{o-erpwn-13}.
ARTC's figure is a conservative estimate that takes into account
infrastructure stress and maintenance requirements and also considers
minimum time separation between trains crossing the single line.
Recall (from Figure~\ref{fig::cperf::congestion}) that, on an
exceptionally busy, day up to 16 trains can arrive at the port. In
this case, single-line utilisation is 44\%. The expected figure is
even less; 31-39\% on weekdays and 19-22\% on weekends. It is easy to
see from this simple analysis that the single-line track is not a
current operational bottleneck and any infrastructural investments at
Port Botany should be directed elsewhere (e.g., increasing the
capacity of rail terminals which are saturated long before the
single-line track, as we show in the next section).

We believe the single-line track is regarded as a bottleneck because it can
become over-subscribed for short periods each day.  In particular, train
arrivals and departures tend to be clustered immediately before and
immediately after the curfew period imposed on freight trains crossing
Sydney's passenger rail network (see Figure~\ref{fig::cperf::arrdep}). Again,
an improved scheduling system for port trains would result in better
utilisation of this section and help mitigate the perception of congestion.

\section{Peak Capacity (Simulation and Predictive Analytics)}
\label{sec::peak}
In this section, we use a simulation-based approach to investigate the
container-carrying capacity of rail infrastructure in the
Enfield-Botany rail corridor.  The outputs from such an analysis,
together with forecasted growth in container volumes, will allow us to
clearly see when the port is likely to become saturated and will help
us identify which pieces of infrastructure will be impacted first.
Such information is necessary to guide long-term infrastructural
planning and to prioritise investment at the port.

Our simulation model is described in
Section~\ref{sec::methodology::sim}.  Unless otherwise noted, we use a
common set of ideal but not unrealistic operational parameters.  These
parameters are described in Appendix~\ref{appx::simparams}.  As part
of our analysis, we undertake several operational and infrastructural
case studies. These case studies are:
\begin{itemize}
\item Port Botany with current rail infrastructure.
\item Port Botany with longer sidings at the DP World terminal.
\item Port Botany with a single centralised rail terminal.
\end{itemize}

A key finding from our analysis is that the peak capacity of current
rail infrastructure at Port Botany is as high as 1.780M TEU per annum;
i.e., over six times higher than actual rail volumes in 2011/12 and
well above the levels necessary to meet the NSW Government's 28\% rail
mode share target for the year 2020 (i.e., 1M TEU).

\subsection{Capacity Case Study 1: Port Botany with current infrastructure}
We developed multiple scenarios to explore the annual peak capacity of rail at
Port Botany.  Scenario 1 (the ``As-Is'' model) represents the current
configuration of the port, i.e.,  there are only two stevedore terminals offering
rail services. Scenario 2 (the ``Soon-to-Be'' model) represents the configuration
of the port in the near future, i.e., all three stevedore terminals 
offer rail services. Table~\ref{table::peak::peak} summarises our main findings.

\begin{table}[t!]
\begin{center}
\begin{tabular}{|l|c|c|c|c|}
\hline
\textbf{Metric} & \multicolumn{2}{c|}{\textbf{Scenario 1: As-Is}} & \multicolumn{2}{c|}{\textbf{Scenario 2: Soon-to-Be}} \\ \hline
Lifts Per Hour & 80\% Current Max & Unchanged & 80\% Current Max & Unchanged \\ \hline
Avg. Trains per day& 16 & 13.11 & 25.4 & 21.3 \\ \hline 
Time (\% Lifting) & 85.8\% & 88.5\% & 86.5\% & 87.7\%  \\ \hline
Time (\% Shunting) & 14.2\% & 11.5\% &  13.5\% & 12.3\% \\ \hline
Peak Volume (TEU) & 1.121M & 0.919M & 1.780M & 1.493M \\ \hline
\end{tabular}
\vspace{-1em}
\caption{Peak capacity at Port Botany. We simulate two scenarios: Scenario 1
represent an As-Is model of the port (with two rail terminals) and Scenario 2
represents a Soon-to-Be model (with three rail terminals). For each scenario,
we present results assuming that: (i) each terminal lifts at a constant 80\% 
of current maximums; (ii) lifting performance at each terminal remains unchanged. }

\label{table::peak::peak}
\end{center}
\end{table}

For the As-Is scenario, we observe that peak capacity of 1.121M TEU
per annum can be reached using two terminals servicing 16 trains per
day on average.  The amount of rail traffic in this case is similar to
the level currently experienced at Port Botany during a ``busy''
weekday.  In the Soon-to-Be case, a peak capacity of 1.780M TEU can be
attained with three terminals servicing 25.4 trains per day, on
average.  For context, under the NSW Government's 28\% rail mode share
target, Port Botany will need to rail 1M TEU per annum by 2020 and
1.4M TEU by 2025.

It is important at this stage to discuss the lift rate performance of
stevedore terminals. Currently lift performance at Port Botany can vary from
day to day and train to train. Chiefly these variations are due to operational
uncertainty; e.g. differences in the number of planned vs actual train
arrivals and differences in the actual vs planned number of containers of each
train.  Such uncertainties can affect the size of the rail crew on the day and
the amount and type of equipment available to service the train.  To model
this variation, we analysed six months of operational data collected from each
terminal operator.  We constructed from the data empirical distributions for
lift rates based on recorded performance stripping and backloading rakes of
length 650m or similar. There are two such distributions, one for each
terminal. We assume that the third operator will be comparable to the current
best observed performance at the port. We used these distributions during
simulation to establish a peak capacity figure for the system under the
assumption that there are no changes to terminal operations in the forseeable
future.

If we assume no operational changes take place at Port Botany's
terminals (i.e., stevedore lift rates are unchanged), the capacity of
the port is 16-18\% lower: 0.919M TEU in the ``As-Is'' scenario and
1.493M TEU for the ``Soon-to-Be''.  In both cases, peak figures are
very close to the stated 28\% rail mode share targets established by
the NSW Government but we can see the system is saturated and
stevedore resources are exhausted.  In order to deal with projected
growth for the year 2020 and beyond, Port Botany will need to
incentivise terminal operators to lift at higher and more consistent
rates than currently. Essentially, this can be achieved by reducing
operational uncertainty (e.g., through better communication in the
rail supply chain) and better allocation of existing resources to
rail.

Our simulation shows that stevedores operating at the port are able to meet
the NSW Government's rail mode share targets for 2020 and beyond with only
changes to operational practices (i.e., consistent lift performance) and no
investment in new infrastructure.  Further our simulation shows that
there are also no bottlenecks or other impediments that prevent the
achievement of these figures and targets (e.g.  single-track or siding
lenghts).

\subsection{Capacity Case Study 2: Longer siding lengths at the DP World terminal}

Within the port community, a range of alternative rake lengths have
also been proposed to maximise rail throughput. For example, because
the DP World rail terminal comprises $3\times350$m rail sidings, it
has been suggested that trains with 350m rakes are optimal with
respect to maximising container volumes. We analysed the impact on
peak capacity from operating container trains with rakes of length
other than 650m.  We explored a range of such alternative
configurations; from rakes that comprise a single wagon (20.3m in
length) to rakes of up to 32 wagons (650m in length). We focus our
discussion on rakes of 17 wagons (350m in length) or longer, as
shorter configurations are less effective.

\begin{figure}[tb]
       \begin{center}
		   \includegraphics[width=0.95\columnwidth]
			{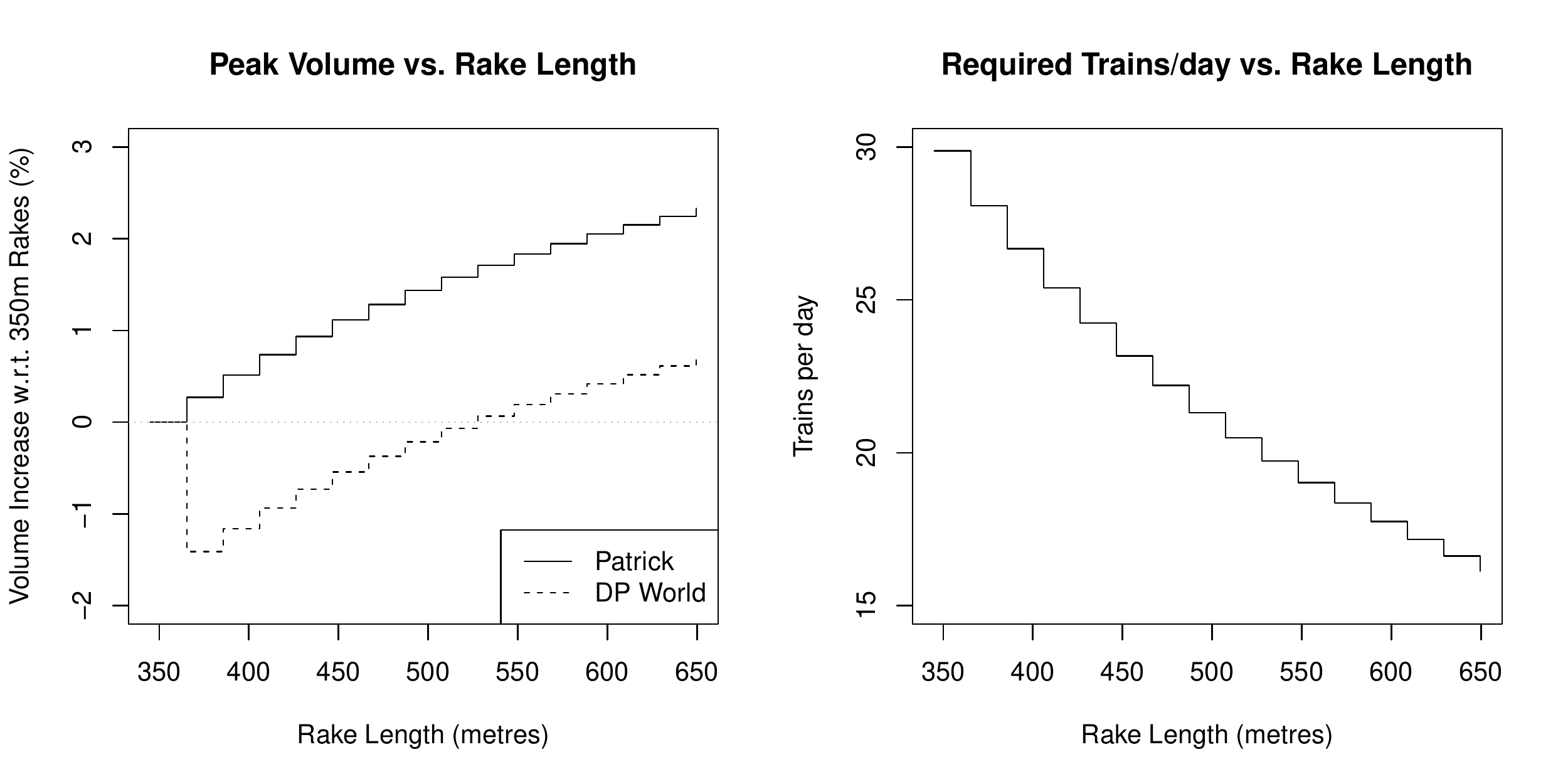}
       \end{center}
        \vspace{-1em}
       \caption{Left: Peak Rail Volume (two stevedores) vs. Rake Length. Values
are indicated as the relative variation w.r.t. the suggested ideal length of
350m to maximise volumes at the DP World terminal.
	Right: Trains per day vs. Rake Length. We observe that maximising volumes 
	with shorter configurations requires a number of trains close to the 
	maximum number of trains that can traverse the single-line section according to
	ARTC (i.e., 36 trains/day).
} 
       \label{fig::rake_len}
\end{figure}

Figure~\ref{fig::rake_len} (left) gives the result of this analysis in terms of
relative container volume variation with respect to the suggested 350m rake
length. We find that 650m 32-wagon rake configurations maximise volume both
across the entire operation and also at each terminal. At the DP World terminal
we can observe that there is less than 1\% difference in volume between 350m
vs. 650m rakes. This observation runs counter to suggestions from industry who
argue that volumes at DP World can be increased by running shorter rakes with
reduced shunting time. Moreover, Figure~\ref{fig::rake_len} (right) shows that
the number of train services per day required to achieve projected volumes
rapidly increases for shorter rake lengths. We observe that, to obtain maximum
volumes with 350m rakes, 30 trains need to access the port daily, while only 16
trains per day are required with 650m rakes.

Our analysis of operational data showed that splitting operations at DP World
introduce only a small overhead to total shunting time. We have also shown, in
our analysis of alternative rake lengths, that operating 350m rakes at the DP
World terminal is marginally less efficient than operating longer 650m rakes. A
logical recommendation would be to extend DP World sidings to accommodate 650m
rakes, aiming at eliminating the shunting and splitting overhead. However, we
quantified the impact of such infrastructure upgrade at 0.4\% increase in peak
rail volume, suggesting that investment should be directed elsewhere in order
to obtain bigger gains.

\subsection{Capacity Case Study 3: A single centralised off-dock rail terminal}
\label{sec::ccs3}
It has recently been suggested that, in order to overcome terminal limitations
and to increase rail capacity at Port Botany, it will be necessary to
consolidate stevedoring operations at a single common-user off-dock rail
terminal \cite{aecom12}. 
Such a scenario has two potentially compelling advantages: (i) improvements in
peak capacity from standardised shunting and lifting operations and (ii) the
elimination of delays for non-dedicated train services that are currently
split between Port Botany's rail terminals. We modelled the proposed facility 
as follows, on advice from Sydney Ports Corporation:
\begin{itemize}
\item There exist three separate rail sidings, each of length 900m.
\item All shunting operations into or out of the terminal are non-conflicting.
\item Shunting operations are streamlined and faster than at current rail
terminals. The proposed improvement is achieved through changes to the current
interface agreement between terminal operators and rail operators. In this
scenario train crews do not disconnect the locomotive from the rake during
servicing and do not need to wait outside the terminal during servicing.
\item All train rakes are 900m in length, carry 136 TEU, and require 88 lifts
to fully strip or load.
\item Trains are serviced at a constant lift rate equal to 80\% of 
current max at Port Botany.
\end{itemize}

\noindent
We find that such a centralised terminal is capable of rail volumes of up to
2.052M TEU per annum; i.e., 15\% higher than the peak rail volume
established in our Soon-to-Be scenario where separate waterfront rail terminals 
are operating at Port Botany.  
It is clear that a centralised terminal appears to provide some gains in peak
rail volumes vs. simply retaining and operating three on-dock rail terminals.
However, if we assume the same consistent lift rates across all three
waterfront terminals in the Soon-to-Be scenario, we find that rail volume
increases up to 1.97M TEU per annum; i.e., a difference of 4\% 
which can be largely attributed to faster shunting times and reduced shunting 
due to longer rakes at the centralised facility. 
Besides offering only marginal gains a single centralised rail terminal has
significant disadvantages:
(i) it requires significant investment in new infrastructure; (ii) it may
require additional resources to ship containers from the facility to the
waterfront. 

Our study shows that a centralised terminal is not necessary to
meet the NSW government's rail mode share targets over the medium term. 
The principal advantages of such a facility (better lift rates,
faster shunting times) appear to be equally achievable through operational
changes and better utilisation of existing rail terminal infrastructure (e.g.
650m dedicated train services, revised interface agreements between rail
operators and stevedores and an emphasis on achieving more consistent lift
rates). In short, operational rather than infrastructural changes seem to be
the key to unlocking the rail capacity of Port Botany .

\section{Conclusions}
\label{sec::conclusions}

In this paper, we analysed the efficacy of rail at Port Botany and
inside the 20km Enfield-Botany rail corridor.  Our objective was
twofold: (i) to evaluate the current performance of rail in and around
the waterfront and; (ii) to investigate the peak rail capacity of both
current and proposed infrastructure and working practices.  In the
course of our study, we analysed six months of operational data
collected at the port over the period September 2012 to February 2013.
We worked with Sydney Ports Coporation and members of the port
community through the Port Botany Landside Improvement Strategy
(PBLIS).

In the first instance, we employed data analytics to evaluate the
current performance of rail resources at Port Botany. We found that
Botany Yard and its associated stevedore terminals appear to be
distinctly under-utilised.  Particular problems that we identified
include under-utilised trains, unproductive staging practices, and
peak-hour congestion stemming from poor train scheduling and
unbalanced allocation of rail resources.

Next, we employed predictive analytics and discrete-event simulation to study
the potential of the rail system. We found that, under a set of ideal (but not
unrealistic) operating conditions, the peak capacity of rail at Port Botany is
1.780M TEU per annum. This result is achieved through operational changes only
and is significantly higher than previously reported estimates
(e.g.~\cite{aecom12}) which assume costly investment in additional
infrastructure. We also investigated the potential of such investments
including: the duplication of the Mascot-Botany line, the extension of rail
sidings at the DP World rail terminal and replacing Port Botany's existing
on-dock rail terminals with a single off-dock centralised facility. In each
instance, we found that new infrastructure is either not necessary or that
investment can be deferred in favour of changes to operational practices. As a
conclusion from our study, particular changes that we recommended are: (i)
replacing fixed servicing windows by dynamic train scheduling; (ii) staging
trains at the Enfield Marshalling Yard instead of Botany Yard; (iii) replacing
non-dedicated non-standard metro/regional traffic with dedicated and standard
650m trains; (iv) the introduction of a minimum rake utilisation policy that
prevents low-volume trains from accessing the port.

Our work focused only on the operational aspects of rail. We have not evaluated
the economic impact of our results nor attempted to construct any cost model.
Our results do show however that Port Botany is well positioned and adequately
equipped to achieve the NSW State Government's stated rail mode share targets
for the year 2020 and beyond. The key to unlocking the potential of Port Botany
over the next decade and beyond lies not in new infrastructure but better
utilisation of existing rail resources. These findings run contrary to
conventional wisdom within the port community, and industry media, which
purports that Port Botany suffers from limited rail resources and requires
additional investment in new infrastructure.
\section{Acknowledgements}

NICTA is funded by the Australian Government through the Department of
Communications and the Australian Research Council through the ICT Centre of
Excellence Program. NICTA is also funded and supported by the Australian
Capital Territory, the New South Wales, Queensland and Victorian Governments,
the Australian National University, the University of New South Wales, the
University of Melbourne, the University of Queensland, the University of
Sydney, Griffith University, Queensland University of Technology, Monash
University, RMIT and Swinburne University.





\begin{thebibliography}{}

\bibitem[\protect\BCAY{{AECOM Australia}}{{AECOM Australia}}{2012}]{aecom12}
{AECOM Australia} \BBOP2012\BBCP.
\newblock {\Bem {S}ydney {P}orts {R}ail {S}trategy {R}eview: {F}inal {R}eport}.
\newblock Consultancy report, commissioned by Sydney Ports Corporation.

\bibitem[\protect\BCAY{Berejiklian\ \BBA\ Gay}{Berejiklian\ \BBA\
  Gay}{2013}]{nsw13}
Berejiklian, G.\BBACOMMA\  \BBA\ Gay, D.\BEDS. \BBOP2013\BBCP.
\newblock {\Bem {NSW} {F}reight and {P}orts {S}trategy}.
\newblock Transport for {NSW}.

\bibitem[\protect\BCAY{Brereton}{Brereton}{2005}]{brereton05}
Brereton, L.\BED. \BBOP2005\BBCP.
\newblock {\Bem {R}ailing {P}ort {B}otany's {C}ontainers: {P}roposals {T}o
  {E}ase {P}ressure {O}n {S}ydney's {R}oads}.
\newblock {NSW Freight Infrastructure Advisory Board}.

\bibitem[\protect\BCAY{Cameron}{Cameron}{2014}]{cameron14}
Cameron, G. \BBOP2014\BBCP.
\newblock \BBOQ Is {M}oorebank running off the rails{?}\BBCQ\
\newblock {\Bem Transport and Logistics News}, {\Bem January 30}.

\bibitem[\protect\BCAY{{{D}epartment of {P}remier and {C}abinet}}{{{D}epartment
  of {P}remier and {C}abinet}}{2011}]{nsw2021}
{{D}epartment of {P}remier and {C}abinet} \BBOP2011\BBCP.
\newblock {\Bem {N}{S}{W} 2021 {A} {P}lan {T}o {M}ake {N}{S}{W} {N}umber
  {O}ne}.

\bibitem[\protect\BCAY{Keating, Cox,\ \BBA\ Krieger}{Keating
  et~al.}{2008}]{ipart-rpblt-08}
Keating, M., Cox, J., \BBA\ Krieger, S. \BBOP2008\BBCP.
\newblock {\Bem Reforming {P}ort {B}otany's Links With Inland Transport:
  {R}eview of the {I}nterface between {L}and {T}ransport {I}ndustries and the
  {S}tevedores at {P}ort {B}otany}.
\newblock Independent Pricing and Regulatory Tribunal of New South Wales.

\bibitem[\protect\BCAY{Lechler\ \BBA\ Page}{Lechler\ \BBA\
  Page}{1999}]{lechler1999desmo}
Lechler, T.\BBACOMMA\  \BBA\ Page, B. \BBOP1999\BBCP.
\newblock \BBOQ {DESMO-J}: An object oriented discrete simulation framework in
  {J}ava\BBCQ\
\newblock In {\Bem Proceedings of the 11th European Simulation Symposium},
  \BPGS\ 46--50.

\bibitem[\protect\BCAY{Ormsby}{Ormsby}{2013}]{o-erpwn-13}
Ormsby, S. \BBOP2013\BBCP.
\newblock \BBOQ {E}nhancing rail projects within the {N}{S}{W} {L}ong {T}erm
  {T}ransport {M}aster {P}lan\BBCQ\
\newblock {\Bem NSW Transport Infrastructure Summit}, {\Bem August}.
\newblock Simon Ormsby, Executive General Manager Strategy \& Growth,
  Australian Rail Track Corporation (ARTC), speaking to industry.

\bibitem[\protect\BCAY{{Parliament of NSW}}{{Parliament of
  NSW}}{2011}]{NSWParl11}
{Parliament of NSW} \BBOP2011\BBCP.
\newblock \BBOQ {E}xamination of proposed expenditure for the portfolio area
  {T}ransport\BBCQ\
\newblock {\Bem {G}eneral {P}urpose {S}tanding {C}ommittee no. 3}, {\Bem 25
  October}.
\newblock Transcript of Meeting.

\end{thebibliography}

\newpage

\appendix
\section{Rail Operations Terminology} 
\label{app::glossary}

\noindent \textbf{Call-out.} Order to proceed shunting a train out terminal
sidings.

\noindent \textbf{Call-up.} Order to proceed shunting a train into terminal
sidings.

\noindent \textbf{Path.} Set of rail resources and time schedule to run a train
between two places in the rail network over a given time-period.

\noindent \textbf{Placement.} Operation consisting of shunting a train into a
siding and detaching the locomotive to start servicing operations within a
terminal.

\noindent \textbf{Rake.} A set of wagons coupled together.

\noindent \textbf{Road.} A railway route connecting two or more places in the
rail network.

\noindent \textbf{Run-around.} Operation consisting of detaching a locomotive
from its train, driving it to the other end of the train and re-attaching it.
This operation is performed before shunting into a terminal.

\noindent \textbf{Shunting.} Operation consisting of moving trains to or from
sidings, as well as dividing trains in sidings. Shunting operations are often
performed at low speeds.

\noindent \textbf{Shuttle.} A train that runs back and forth, usually over a
relatively short distance, between two locations in the rail network.

\noindent \textbf{Siding.} A section of track off the main line. Sidings are
often used for staging or servicing trains.

\noindent \textbf{Staging.} Holding a train, normally at a siding or within a
yard. 

\noindent \textbf{Yard / Marshalling Yard.} A complex set of rail tracks used
for storing, sorting, or loading/unloading, rakes and/or locomotives. A
\textit{marshalling yard} is a type of yard used to separate and rearrange
rakes.


\section{Peak Capacity Simulation Parameters}
\label{appx::simparams}
To evaluate peak capacity we simulate rail operations at the port under
a set of operational parameters which we found to be ideal. Some of these 
parameters have been suggested previously (e.g. ~\cite{brereton05,aecom12}) 
but to the best of our knowledge they have never been analysed holistically.
These parameters are:
\begin{itemize}
\item All trains are dedicated 650m shuttles and visit one Stevedore only.
\item Trains are scheduled dynamically and originate at the Enfield Marshalling Yard.
\item Trains are staged inside Botany Rail Yard prior to servicing.
\item Trains are full when they arrive and fully backloaded when they depart.
\item All train rakes are homogeneous, comprise 32 wagons and have a capacity of 96 TEU.
\item Shunting operations at stevedore terminals proceed without delays.
\item Stevedores operate 24/7 and lift at a constant rate equal to 80\% of 
current max performance.
\end{itemize}



\end{document}